\documentclass[twocolumn,twocolappendix]{aastex631}
\usepackage[T1]{fontenc}
\usepackage{amsmath,mathtools,mathrsfs,amssymb}
\usepackage{hyperref}
\usepackage{verbatim}
\usepackage{xspace}
\usepackage{graphicx}
\usepackage{longtable}

\newcommand{\bhspin}{a_*}
\newcommand{\rhigh}{R_{\mathrm{high}}}
\newcommand{\rlow}{R_{\mathrm{low}}}
\newcommand{\sgra}{{Sgr\,A*}}
\newcommand{\kB}{k_{\rm B}}
\newcommand{\mnet}{\left|m\right|_{\rm net}}
\newcommand{\vnet}{\left|v\right|_{\rm net}}
\newcommand{\avgm}{\left<\left|m\right|\right>}

\newcommand{\iharm}[0]{{\tt{iharm3d}}}
\newcommand{\ipole}[0]{{\tt{ipole}}}
\newcommand{\igrmonty}[0]{{\tt{igrmonty}}}

\begin{document}

\shorttitle{Hydrogen vs.~Helium in Black Hole Observables}

\title{Effects of Hydrogen vs.~Helium on Electromagnetic Black Hole Observables}

\shortauthors{Wong \& Gammie}

\correspondingauthor{George~N.~Wong}
\email{gnwong@ias.edu}

\author[0000-0001-6952-2147]{George~N.~Wong}
\affiliation{School of Natural Sciences, Institute for Advanced Study, 1 Einstein Drive, Princeton, NJ 08540, USA}
\affiliation{Princeton Gravity Initiative, Princeton University, Princeton, New Jersey 08544, USA}

\author[0000-0001-7451-8935]{Charles~F.~Gammie}
\affiliation{Department of Physics, University of Illinois, 1110 West Green Street, Urbana, IL 61801, USA}
\altaffiliation{Illinois Center for Advanced Study of the Universe, 1110 West Green Street, Urbana, IL 61801, USA}
\altaffiliation{Department of Astronomy, University of Illinois, 1002 West Green Street, Urbana, IL 61801, USA}
\altaffiliation{National Center for Supercomputing Applications, 1205 West Clark Street, Urbana, IL 61801, USA}

\begin{abstract}
The centers of our galaxy and the nearby Messier 87 are known to contain supermassive black holes, which support accretion flows that radiate across the electromagnetic spectrum. Although the composition of the accreting gas is unknown, it is likely a mix of ionized hydrogen and helium. We use a simple analytic model and a suite of numerical general relativistic magnetohydrodynamic accretion simulations to study how polarimetric images and spectral energy distributions of the source are influenced by the hydrogen/helium content of the accreting matter. We aim to identify general trends rather than make quantitatively precise predictions, since it is not possible to fully explore the parameter space of accretion models. If the ion-to-electron temperature ratio is fixed, then increasing the helium fraction increases the gas temperature; to match the observational flux density constraints, the number density of electrons and magnetic field strengths must therefore decrease. In our numerical simulations, emission shifts from regions of low to high plasma $\beta$---both altering the morphology of the image and decreasing the variability of the light curve---especially in strongly magnetized models with emission close to the midplane. In polarized images, we find that the model gas composition influences the degree to which linear polarization is (de)scrambled and therefore affects estimates for the resolved linear polarization fraction. We also find that the spectra of helium-composition flows peak at higher frequencies and exhibit higher luminosities. We conclude that gas composition may play an important role in predictive models for black hole accretion.
\end{abstract}


\section{Introduction}

\nocite{EHTC_2019_1}
\nocite{EHTC_2019_4}
\nocite{EHTC_2019_5}
\nocite{EHTC_2019_6}

The accretion flows around the black holes at the center of our galaxy, hereafter \sgra, and the center of the nearby elliptical galaxy Messier 87, hereafter M87, are typically modeled as radiative inefficient accretion flows (RIAFs), which comprise geometrically thick disks of infalling plasma \citep[e.g.,][]{ichimaru_1977_BimodalBehaviorAccretion,rees_1982_IonsupportedToriOrigin,narayan_1995_ADAF,reynolds_1996_QuiescentBlackHole}.  RIAF flows are Coulomb collisionless, but kinetic plasma instabilities may enable particle-wave interactions to mediate the ion and electron distribution functions so that they can be treated as a fluid with well defined temperatures (see \citealt{kunz_2014_FirehoseMirrorInstabilities} and discussion therein). Given a model for the particle distribution functions, the total internal energy of the fluid, and a prescription for the temperature ratio between the ions and the electrons, it is then possible to compute the ion and electron temperatures. For the parameters relevant for the accretion flows around M87 and \sgra, it is likely that the electron temperature differs from the ion temperature \citep[e.g.,][]{shapiro_1976_twotemp,mahadevan_1997_AreParticlesAdvectiondominated,quataert_1998_ParticleHeatingAlfvenic,sadowski_2017_RadiativeTwotemperatureSimulations,chael_2018_RoleElectronHeating,ryan_2018_TwotemperatureGRRMHDSimulations}.

The Event Horizon Telescope (EHT) collaboration has released horizon-scale images of the putative supermassive black holes, M87 and \sgra. In its current configuration, the EHT interferometer reconstructs images at an operational frequency of approximately $230\,$GHz; for M87- and \sgra-like accretion systems, radiation at $230\,$GHz is dominated by synchrotron emission (see \citealt{yuan_2014_HotAccretionFlows}), whereby electrons emit polarized radiation as they spiral around magnetic field lines. The frequency-dependent amount of synchrotron emission is controlled by three parameters: the local magnetic field strength and orientation, the number density of the emitting electrons, and the electron distribution function.

General relativistic magnetohydrodynamics (GRMHD) simulations are widely used to study the environment near the event horizon via numerical simulation of the time-dependent behavior of the plasma flow as it accretes \citep[see especially][and also the wind-fed models of the galacic center from \citealt{ressler_2018_HydrodynamicSimulationsInner,ressler_2020_InitioHorizonscaleSimulations}, which include a part-helium gas]{EHTC_2019_5,EHTC_2021_8,EHTC_2022_5}.
The output of the simulations can then be processed with general relativistic ray tracing (GRRT) codes to generate simulated observables, like images and spectra. GRMHD simulations typically output the total rest-mass density and specific internal energy of the fluid as a function of space and time. These quantities are translated into the $n_e$ and electron distribution function required for GRRT by adopting models for the composition of the gas and the partitioning of energy between the constituent electrons and ions of the plasma.

The composition of the accreting material is determined by the sources that feed the accretion flow. In the case of supermassive black holes, the flow is likely fed by gas from clouds and stellar winds near the center of a galaxy. If this gas has approximately solar composition then it is hydrogen rich and we might expect the helium mass fraction to be $\approx 1/4$. The composition of gas within the central few kiloparsecs of M87 has been considered, but it is not well constrained. Photometric measurements infer of order a few times solar metallicity and find that some globular clusters may be extremely helium-rich (see, e.g., \citealt{montes_2014_AgeMetallicityGradients} and \citealt{bellini_2015_UVInsightsComplex}). \citet{martins_2007_StellarWindProperties} modeled the spectra of stars within a parsec of the galactic center and found evidence for a low H/He ratio, which implies that a significant fraction of the accreting matter could be helium, if accretion is due to stellar winds (see also \citealt{cuadra_2008_VariableAccretionEmission,calderon_2016_ClumpFormationColliding}). Nevertheless, the galactic center also hosts a large amount of molecular and ionized hydrogen gas, which could also supply mass to the \sgra~accretion system.

In this paper, we explore how the composition of the emitting plasma affects properties of electromagnetic observables for the M87 and \sgra~RIAF systems targeted by the EHT. Measurements of the horizon-scale flux density in these systems infer large brightness temperatures in excess of $10^9\,$K. The ionization temperatures for both hydrogen and helium are well below this value: hydrogen is fully ionized at $T \gtrsim 10^4\,$K and helium is fully ionized at $T \gtrsim 2.9 \times10^4\,$K. We therefore assume that the plasma comprises free electrons and ions, with pure protons for hydrogen and nuclei with two protons and two neutrons for helium.

Most models consider only a pure hydrogen plasma. Here we consider a hydrogen/helium plasma with no metal ions. The composition of the gas may affect the fraction of energy dissipated into electrons and ions, since the ion Larmor radius is four times larger for helium than hydrogen. The composition may also affect the plasma dynamics in regions with $T_e \sim T_i$, since electron cooling potentially removes $2/3$ rather than just $1/2$ of the gas pressure. We will not consider these effects, but instead illustrate the importance of composition with a simple model in which the ions and electrons are assumed thermal and $R \equiv T_i/T_e$ is assigned using the so-called $\rhigh$ model of \citet{moscibrodzka_2016_rhigh}.

This paper is organized as follows:
We describe a simple one-zone model in Section~\ref{sec:onezone} and use it to investigate the effect of varying between pure hydrogen and pure helium gas composition. In Section~\ref{sec:numericalmethods}, we first describe the details of the numerical modeling procedure we use to synthesize polarimetric images and spectral energy distributions and then motivate the parameter space for our models. We report the results of the numerical exploration in Section~\ref{sec:grmhdresults} and conclude in Section~\ref{sec:conclusion}.

\section{One Zone Model}
\label{sec:onezone}

We first test how gas composition influences the properties of simple one-zone emission models that have been tuned to M87 and \sgra. In the one-zone model, we treat the source as a spherical ball in flat space with a radius $R_0 \sim$ comparable to the observed source size (see \citealt{EHTC_2019_5,EHTC_2022_5} for more detail) and with uniform particle number density, temperature, and magnetic field. We tune the model parameters until the flux density produced by the model at $230\,$GHz matches the observed flux density.

Our one-zone model makes the following assumptions:
\begin{itemize}
    \item the source is compact with $R_0 = 5\,GM/c^2$,
    \item the gas pressure $\sim$ the magnetic pressure ($\beta = 1$),
    \item the relatively low collisionality of the flow causes the ions to be preferentially heated, such that $T_i = 3 T_e$,
    \item the ions are slightly subvirial so that the ions are nonrelativistic and the electrons are moderately relativistic.
\end{itemize}

Under these assumptions, the $230\,$GHz emission is produced primarily by the synchrotron process \citep{yuan_2014_HotAccretionFlows}, in which electrons radiate as they circle around magnetic field lines. The overall emission, absorption, and rotation radiative transfer coefficients are computed by integrating over a distribution of electrons with different momenta. We assume that the electrons occupy a relativistic, thermal distribution function; emission at frequencies relevant to EHT-like observations likely sample the core of the distribution function, so any higher-frequency non-thermal components are less likely to influence the results.
We require that the total emission from our one-zone model reproduce the observed flux density by self-consistently altering the electron number density until the integrated emission across the sphere matches a target value.

The $\beta = 1$ requirement allows us to write the magnetic field strength as a function of electron temperature and number density
\begin{align}
    B^2 = 8 \pi \left( n_i \kB T_i + n_e \kB T_e \right),
\end{align}
where $n_i$ and $n_e$ are the total numbers of ions and electrons in the gas.
If a fraction $F$ of all of the ions in the system are helium, then the numbers of hydrogen and helium ions are
\begin{align}
    n_{i,\rm H} &= \dfrac{1 - F}{1 + F} n_e, \\
    n_{i,\rm He} &= \dfrac{F}{1 + F} n_e.
\end{align}
Note that helium abundance is often written in terms of the helium mass fraction $Y$. When the mass fraction of all elements heavier than helium $Z \ll Y$, $F$ is related to $Y$ by
\begin{align}
    F = \dfrac{Y}{4 - 3Y}.
\end{align}

The temperature of the electrons can be written in terms of the ratio of the fluid internal energy to its rest-mass density $u/\rho$ by simultaneously choosing an ion-to-electron temperature ratio $R$ and requiring that the sum of the electron and ion energies equal the internal energy of the fluid $u = u_e + u_i$. The dimensionless electron temperature is then
\begin{align}
    \Theta_e \equiv& \dfrac{T_e \, \kB}{m_e c^2} \nonumber \\ 
    =& \dfrac{u}{\rho c^2} \times \left( \dfrac{m_p}{m_e} \, \dfrac{ \left( \gamma_e - 1\right) \left(\gamma_i -1\right)}{y \left( \gamma_i-1\right)  + z \left(\gamma_e - 1 \right) R} \right),
    \label{eqn:thetae_prescription}
\end{align}
where $\gamma_e$ and $\gamma_i$ are the adiabatic indices of the electron and ion fluids, $m_p$ and $m_e$ are the mass of a proton and an electron, and
\begin{align}
    y &= \dfrac{1 + F}{1+3F}, \\
    z &= \dfrac{1}{1+3F},
\end{align}
so that $1/y$ and $1/z$ are the number of electrons and nucleons per unionized atom respectively. Then the electron temperature is
\begin{align}
\label{eqn:rhighdiff}
    T_e =
    \dfrac{m_p}{\kB} \dfrac{u}{\rho} \dfrac{\left(\gamma_e - 1\right)\left(\gamma_i - 1\right)\left(1 + 3F\right)}{R\left(\gamma_e -1 \right) + \left(\gamma_i - 1\right) \left( 1 + F\right)}.
\end{align}

Given values for $F$ and $u/\rho$, we compute the local synchrotron emissivity and absorptivity using the fitting formul\ae{} of \citet{leung_2011_NumericalCalculationMagnetobremsstrahlung} assuming that the pitch angle between the line of sight and the magnetic field is $\pi / 3$. This procedure results in a nonlinear equation for the total electron number density, which we solve numerically for different values of $F$.

\begin{figure}[t!]
\centering
\includegraphics[trim=0.2cm 0cm 0.2cm 0cm,clip,width=1\linewidth]{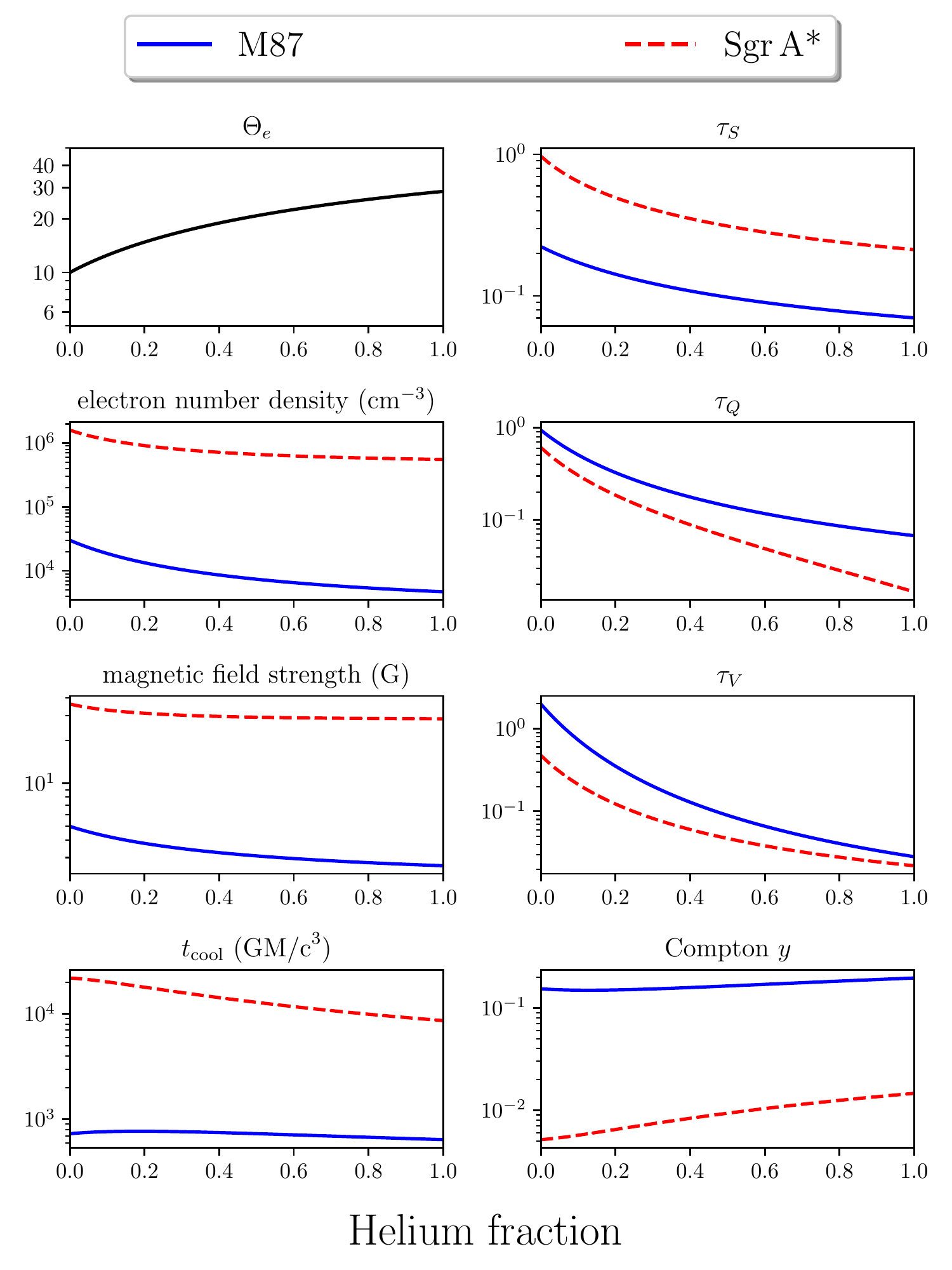}
\caption{Parameters and output of one-zone models that reproduce observed flux density at $230\,$GHz for both M87 and \sgra, plotted vs.~helium fraction $F$ (fraction of ions that are helium). Here, $t_{\rm cool}$ is the synchrotron cooling time, and $\tau_S, \tau_Q$, and $\tau_V$ are the optical depths to absorption and Faraday conversion and rotation, respectively. Models are initialized for a gas with total specific internal energy such that the electron temperature $\Theta_e = 10$ for a pure hydrogen with uniform $T_i/T_e = 3$.}
\label{fig:onezone}
\end{figure}

Figure~\ref{fig:onezone} shows how the M87 and \sgra~one-zone models change as $F$ is varied between $0-100\,\%$ for a reference value of $u/\rho$ that yields $\Theta_e = 10$ for the $F=0$ hydrogen gas. In general, we find that the number density of electrons $n_e$ and the magnetic field strength $B$ decrease as the electron temperature $\Theta_e$ increases. Recall that even though the mass density, internal energy, and magnetic field strength change, the plasma $\beta$ parameter and the magnetization $\sigma \equiv b^2 / \rho$ do not. This choice ensures that altering the gas composition does not change the solution to the evolution equations and enables a sort of apples-to-apples comparison. We discuss this point in greater detail in the context of conservation equations in Section~\ref{sec:fluidevolution}.

We use these results to compute optical depths to different radiative transfer processes, including absorption $\tau_S$ as well as Faraday conversion $\tau_Q$ and Faraday rotation $\tau_V$. We find that in all cases, the depths decrease as $F$ increases and $n_e$ and $B$ decrease. We also compute the synchrotron cooling time $t_{\rm cool}$ for the models and find that it remains large, i.e., $t_{\rm cool} \gg t_{\rm infall}$. In the case of \sgra, we find that the relative importance of Comptonization, gauged by the Compton $y$ parameter, remains small, while for M87 it remains $\approx 0.1$, regardless of the gas composition.

\section{Numerical Models}
\label{sec:numericalmethods}

One-zone models provide useful intuition, however, they are too simple for detailed comparison to observations, since they do not in general reproduce the image morphology or variability statistics observed in sources like M87 and \sgra. We thus validate the results of the one-zone model with a set of simulated images and spectra of RIAF accretion flows in the Kerr spacetime produced with the {\tt{}PATOKA} pipeline \citep{wong_2022_PATOKA}, in which GRMHD simulations are ray-traced to produce simulated electromagnetic observables. 

The fluid simulations are produced with \iharm~\citep{gammie_2003_harm,prather_2021_iharm}, the polarimetric images are produced with \ipole~\citep{moscibrodzka_2018_ipole}, and the spectra are produced with \igrmonty~\citep{dolence_2009_grmonty}. The results of GRMHD simulations are compromised by numerical floors in regions with high $\sigma$ such as the jet region around the poles.  The plasma density in these regions is unphysically high due to the floors. We therefore introduce a $\sigma$ cutoff when generating images and spectra, setting the plasma density to zero in regions with $\sigma > 1$.

\subsection{Fluid evolution}
\label{sec:fluidevolution}

We generate realizations of the fluid flow by solving the equations of general relativistic magnetohydrodynamics (GRMHD), which take the form of a hyperbolic system of conservation laws
\begin{align}
\partial_t \left( \sqrt{-g} \rho u^t \right) &= -\partial_i \left( \sqrt{-g} \rho u^i \right), \label{eqn:massConservation}\\
    \partial_t \left( \sqrt{-g} {T^t}_{\nu} \right) &= - \partial_i \left( \sqrt{-g} {T^i}_{\nu} \right) + \sqrt{-g} {T^{\kappa}}_{\lambda} {\Gamma^{\lambda}}_{\nu\kappa},  \\
\partial_t \left( \sqrt{-g} B^i \right) &= - \partial_j \left[ \sqrt{-g} \left( b^j u^i - b^i u^j \right) \right], \label{eqn:fluxConservation}
\end{align}
with the constraint
\begin{align}
\partial_i \left( \sqrt{-g} B^i \right) &= 0. \label{eqn:monopoleConstraint} 
\end{align}
Here, the plasma rest mass density is $\rho_0$ and its four-velocity is $u^\mu$. The magnetic field is represented by the $b^\mu$ four-vector, and the stress--energy of the fluid is denoted by the rank-2 tensor $T^{\mu\nu}$. The spacetime geometry enters through the metric $g_{\mu\nu}$ and its derivatives via the Christoffel symbol $\Gamma$ and its determinant $g$. More information can be found in \citet{wong_2022_PATOKA}.

The stress--energy tensor $T^{\mu\nu}$ contains contributions from both the fluid and the electromagnetic field
\begin{align}
{T^{\mu\nu}} &= \left( \rho + u + P + b^{\lambda}b_{\lambda}\right)u^{\mu}u^{\nu} \nonumber  \\
&\qquad \quad + \left(P + \frac{b^{\lambda}b_{\lambda}}{2} \right){g^{\mu}}^{\nu} - b^{\mu}b^{\nu},
\label{eqn:mhdTensor}
\end{align}
where $u$ is the internal energy of the fluid and the fluid pressure $P$ is related to its internal energy through a constant adiabatic index $\hat{\gamma}$ with $P = \left(\hat{\gamma} - 1\right) u$.

We have thus far not specified the gas composition. Notice that the evolution equations are invariant under rescalings of the stress--energy tensor by an arbitrary factor $\mathcal{M}$, so that different values of, e.g., $\rho$ may correspond to the same morphological fluid evolution as long as the other fluid parameters are scaled appropriately. Thus, any particular solution to the evolution equations is a member of a congruence class of particular solutions. 

Each congruence class can then be represented by a set of spacetime fields for the normalized rest-mass density, the two ratios of the rescalable quantities $\beta$ and $\sigma$, the magnetic field direction, and the fluid velocity
\begin{align}
    S =  \begin{pmatrix} 
    \rho(t;r,\theta,\phi) \\
    \beta(t;r,\theta,\phi) \\ 
    \sigma(t;r,\theta,\phi) \\
    \hat{B}(t;r,\theta,\phi) \\
    u^i(t;r,\theta,\phi)
    \end{pmatrix}
\end{align}
along with the equivalence relation
\begin{align}
    \rho \sim \mathcal{M} \rho, \quad \mathcal{M} \in \mathbb{R}^{+} ,
\end{align}
which leaves the values of the latter seven fields unchanged.

In this representation, fitting a simulation to a target observational flux density corresponds to selecting an element from the congruence class. In the one-zone model, we require $\beta = 1$ and $\Theta_e = 10$ for pure hydrogen; in the following sections our congruence classes are the output of the GRMHD simulations. We thus perform an apples-to-apples comparison by fixing the congruence class (selecting a particular fluid snapshot) and selecting different members (setting gas composition and fitting the density scale).

\subsection{Model space}

Our models live in a high-dimensional space, which spans black hole mass, accretion rate, different configurations of the accretion flow magnetic field, and different black hole angular momenta (magnitude and orientations relative to both the accretion and the line of sight). Theoretical modeling uncertainties introduce another functional degree of freedom in thermodynamics of the radiating electrons, as the energization, cooling, and coupling mechanisms for the ions and electrons are yet underspecified. In this paper, we consider a subset of the full parameter space relevant to EHT-like sources.

The magnetization of the accretion flow can be qualitatively divided into two states according to the relative magnitude of the magnetic pressure compared to the fluid ram pressure near the horizon. When the two pressures are comparable, the flow enters the so-called magnetically arrested disk (MAD; \citealt{bisnovatyi-kogan_1974_AccretionMatterMAD,igumenshchev_2003_ThreedimensionalMagnetohydrodynamicSimulations,narayan_2003_MagneticallyArrestedDisk}) state, which contrasts with standard and normal evolution (SANE; \citealt{narayan_2012_sane,sadowski_2013_EnergyMomentumMass}). SANE flows are turbulent but relatively steady and disk-like. In the near-horizon region of MAD models, large tubes of magnetic flux arrest the inward motion of the flow, and accretion instead proceeds in chaotic, isolated bursts mediated by transient filaments of hot plasma that thread the region between the hole and the bulk of the accreting material at large radius. Although recent observations suggest a MAD state for both M87 and \sgra~\citep{EHTC_2019_5,EHTC_2021_8,EHTC_2022_5}, we consider simulations of both SANE and MAD flows.

The black hole angular momentum is typically expressed in terms of a dimensionless spin parameter $\bhspin \equiv J c / GM^2$ with $\left| \bhspin \right| \le 1$, where $J$ is the magnitude of the angular momentum of the black hole and $M$ is its mass. Negative values of $\bhspin$ correspond to the scenario where the angular momentum of the accretion flow and the spin of the black hole are anti-aligned. There is no \emph{a priori} reason to assume that the angular momenta of the hole and the flow are aligned or anti-aligned; the angle between the two axes is known as the tilt of the system and is bounded by $90^\circ$. In this paper, we only consider systems with zero tilt but cover five different black hole spins $\bhspin = -15/16, -1/2, 0, 1/2,$ and $15/16$.

The magnetization and black hole spin are the only two parameters that must be specified for the GRMHD simulation, since: (1) the equations of GRMHD are invariant under rescalings of both the black hole mass and the accretion rate; (2) the equations of GRMHD evolve the total energy of the fluid (and are thus $\sim$ agnostic to the microphysical thermodynamics); and (3) the GRMHD simulations are three-dimensional, so the inclination angle $i$ between the spin of the black hole and the line of sight can be varied after the fact.

Since we consider only the M87 and \sgra~systems, we set the mass of the black hole directly according to previous measurements. Furthermore, we can fix the accretion rate by enforcing the requirement that the simulated flux density at the $230\,$GHz operational frequency of the EHT match the observed flux density, since increasing the accretion rate increases the observed flux density (see Appendix D of \citealt{wong_2022_PATOKA} for caveats and more detail). The values we choose for $M$ and $F_{230\,\rm GHz}$ are discussed in Appendix~\ref{app:paramestimate}.

In the case of M87, there is clear evidence of a large-scale jet (see \citealt{walker_2018_m87jet}). In the systems we consider with no tilt, the jet is launched along the symmetry axis of the system, so we set the inclination angle to either $17^\circ$ or $163^\circ$ to be consistent with the large-scale jet orientation and the observed brightness asymmetry seen in \citet{EHTC_2019_5}. It is not yet clear whether \sgra~hosts a jet, but we do not have clear observational constraints on a jet orientation, which could constrain our freedom in setting the inclination angle. Recent results from both GRAVITY \citep{gravitycollaboration_2018_DetectionOrbitalMotions} and the EHT \citep{EHTC_2022_5} infer that we observe \sgra~at a low inclination angle; however, we remain agnostic to this prediction to allow for deviations and model similar systems, which may in general be oriented at arbitrary inclinations to Earth. We thus consider three inclinations, including face-on at $10^\circ$, edge-on at $90^\circ$, and an intermediate case at $50^\circ$. 

There is a more general freedom in assigning the electron distribution function from the local total internal energy output by the fluid simulation. In the M87 and \sgra~RIAF systems, radio frequency emission is presumed to be produced by the synchrotron process, and at the plasma parameters relevant for these systems, the radio frequency emission is likely produced near the thermal core of the distribution function. We thus assume the electron population is well modeled as a relativistic thermal Maxwell--J\"{u}ttner distribution. We determine the local ion-to-electron temperature ratio of Equation~\ref{eqn:thetae_prescription} following the prescription described in 
\citet{moscibrodzka_2016_rhigh}, which is motivated by models for electron heating in a turbulent collisionless plasma that preferentially heats the ions when the gas pressure exceeds the magnetic pressure.
In this model, $R$ is written in terms of the local fluid plasma $\beta$ ratio of gas to magnetic pressures, $\beta \equiv P_{\mathrm{gas}}/P_{\mathrm{mag}}$:
\begin{align}
    R = \dfrac{ \rlow + \rhigh \, \beta_{\mathrm{R}}^2 } { 1 + \beta_{\mathrm{R}}^2 },
    \label{eqn:rhigh}
\end{align}
where $\beta_{\mathrm{R}} \equiv \beta / \beta_{\mathrm{crit}}$, and $\beta_{\mathrm{crit}}$, $\rlow$, and $\rhigh$ are parameters that control the temperature ratio in regions of low (high) $\beta$ where the plasma is dominated by gas (magnetic) pressure. In this work, we set $\rlow = 1$ and $\beta_{\rm crit} = 1$.

\section{Results from GRMHD}
\label{sec:grmhdresults}

We now use a set of GRMHD simulations to study the effect of gas composition on polarized image morphology at $86$, $230$, and $345\,$GHz, light curve variability at $230\,$GHz, and the spectral energy distribution. To limit the dimensionality of the problem, we only consider the two limiting cases with either pure hydrogen or pure helium gas, although real systems may lie between these two extremes. Our procedure involves taking a particular fluid snapshot from a GRMHD simulation and finding $\mathcal{M}$ for each of the hydrogen and helium cases in order to reproduce the target flux density. In all models, the mass accretion rate is lower for the helium scenario. The GRMHD simulations used in this analysis were generated with the assumption that the adiabatic index of the fluid is constant, independent of position, time, $\rhigh$, and gas composition.\footnote{This assumption is technically inconsistent with the post processing; however, the inconsistencies it introduces will be small relative to the sense of the results.} More detail about the simulation procedure can be found in \citealt{wong_2022_PATOKA}.

\begin{figure*}
\centering
\includegraphics[trim=0cm 0cm 0cm 0cm,clip,width=1\linewidth]{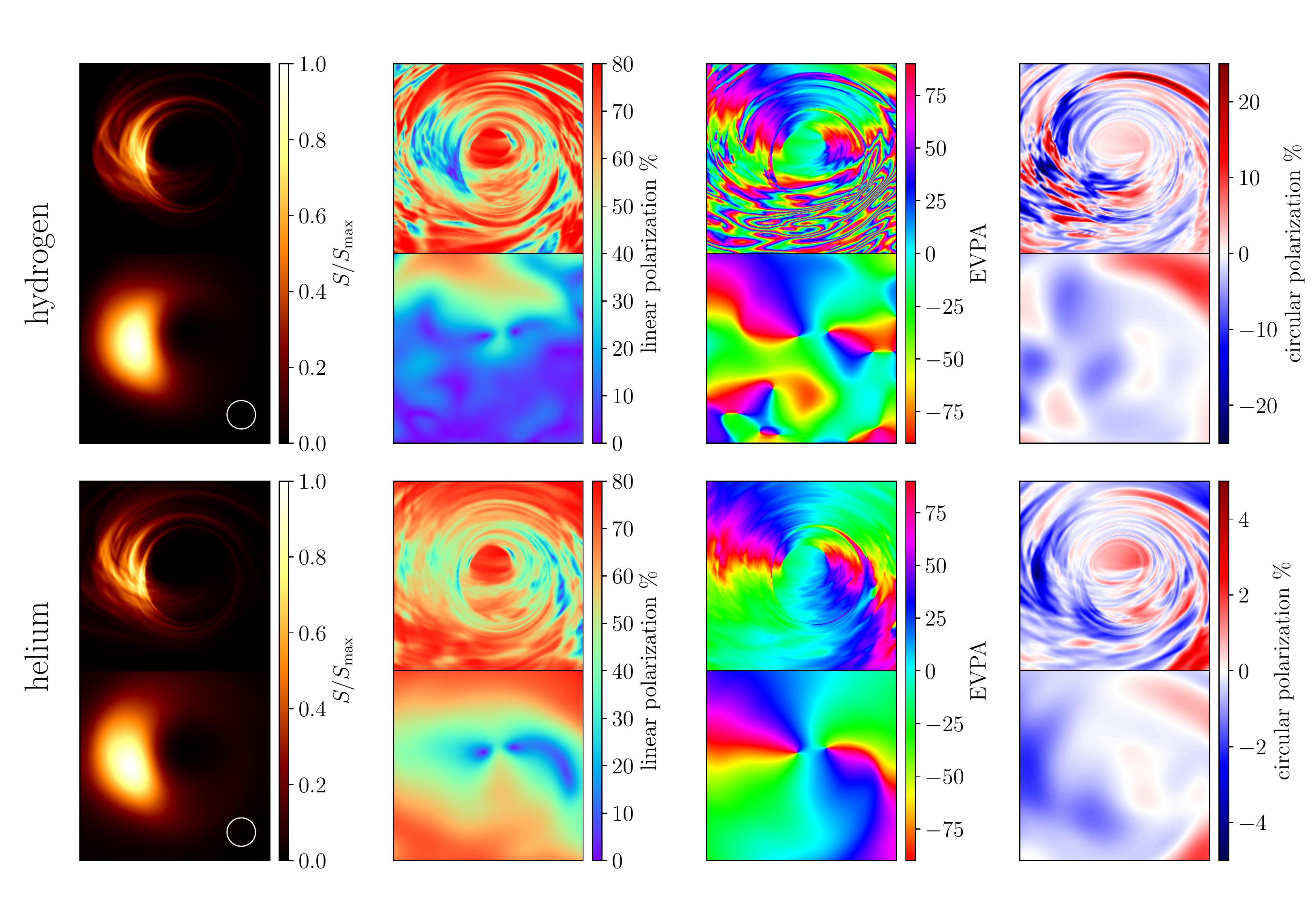}
\caption{Comparison of the same fluid snapshot from a \sgra-like SANE $\bhspin = 0.94$ model at $50^\circ$ inclination with $\rhigh = 10$ at $230\,$GHz for hydrogen vs.~helium gas composition with and without $15\,\mu$as blur (white circle) to simulate the effects of a finite beam size. Columns show total intensity, linear polarization fraction, electric vector position angle (EVPA), and circular polarization fraction. In the hydrogen model, the EVPA varies rapidly in the lower part of the image due to Faraday effects, so the blurred linear polarization fraction is significantly decreased compared to the helium case, where the EVPA is coherent over scales $\sim$ the beamsize. 
}
\label{fig:imgsblur}
\end{figure*}

Figure~\ref{fig:imgsblur} shows a set of polarimetric images produced for an example fluid snapshot at $230\,$GHz and shown with and without a $15\,\mu$as Gaussian blurring kernel applied, to simulate a finite-resolution observation of a \sgra-like model. Each row shows the total intensity $S$, linear polarization fraction, electric vector position angle (EVPA), and circular polarization fraction. Although the total intensity image looks very similar for the pure-hydrogen and pure-helium cases, the polarimetric properties change significantly. In this case, the denser gas in the hydrogen images scrambles the EVPA in the lower part of the image. When blurred, the rapidly varying EVPA averages to a lower overall linear polarization: the hydrogen image is dominated by a sub-$20\%$ linear polarization compared to $40-60\%$ for the helium image.

Many of the numerical models reproduce the trends seen in Figure~\ref{fig:imgsblur} and reproduce the one-zone model expectation that $n_e$ decreases from the pure-hydrogen case to the pure-helium one. However, some do not: SANE flows with large $\rhigh$, and particularly those with negative spin, exhibit trends in the opposite sense as the gas composition is varied. This difference is primarily due to how the structure of the emission region changes between the hydrogen and helium scenarios. Even though the total mass of the system is lower when the gas is pure helium, the primary emission location can shift between the hydrogen and helium scenarios and move from a low-density gas (e.g., in the jet funnel wall in the hydrogen case) to high-density one (e.g., in the midplane for the helium case).

We now discuss the structure of emission before considering how polarimetric images, the variability at $230\,$GHz, and spectra produced by the system change between the two different gas compositions. In this section, we only identify and describe broad, qualitative trends, since our models do not uniformly cover parameter space. We leave a detailed, quantitative study to future work.

\subsection{Emission structure}

\begin{figure}
\centering
\includegraphics[width=1
\linewidth]{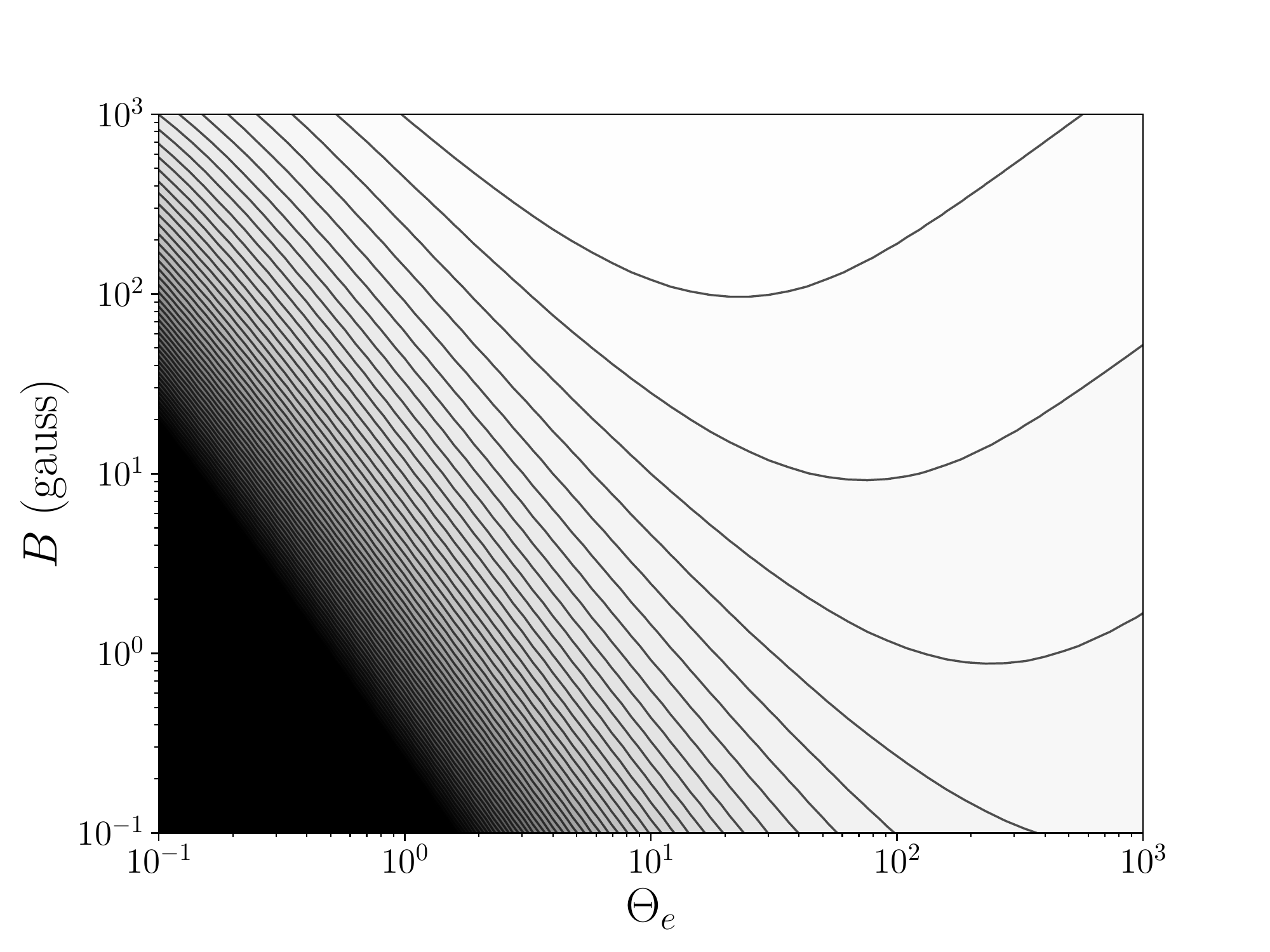}
\caption{Contours of log$_{10}$ per-electron synchrotron emissivity at fixed frequency $\nu = 230\,$GHz and pitch angle $\theta = \pi/3$ across electron temperature $\Theta_e$ and local magnetic field strength $B$ in ranges relevant to RIAF accretion flows near M87 and \sgra. Altering the gas composition from hydrogen to helium changes the magnetic field strength by a uniform factor, but the local temperature change can differ for different locations across the domain. Thus, the ratio of local emissivities for hydrogen vs.~helium differs across the flow.
}
\label{fig:emissivity_contour}
\end{figure}

\begin{figure*}[t!]
\centering
\includegraphics[width=1\linewidth]{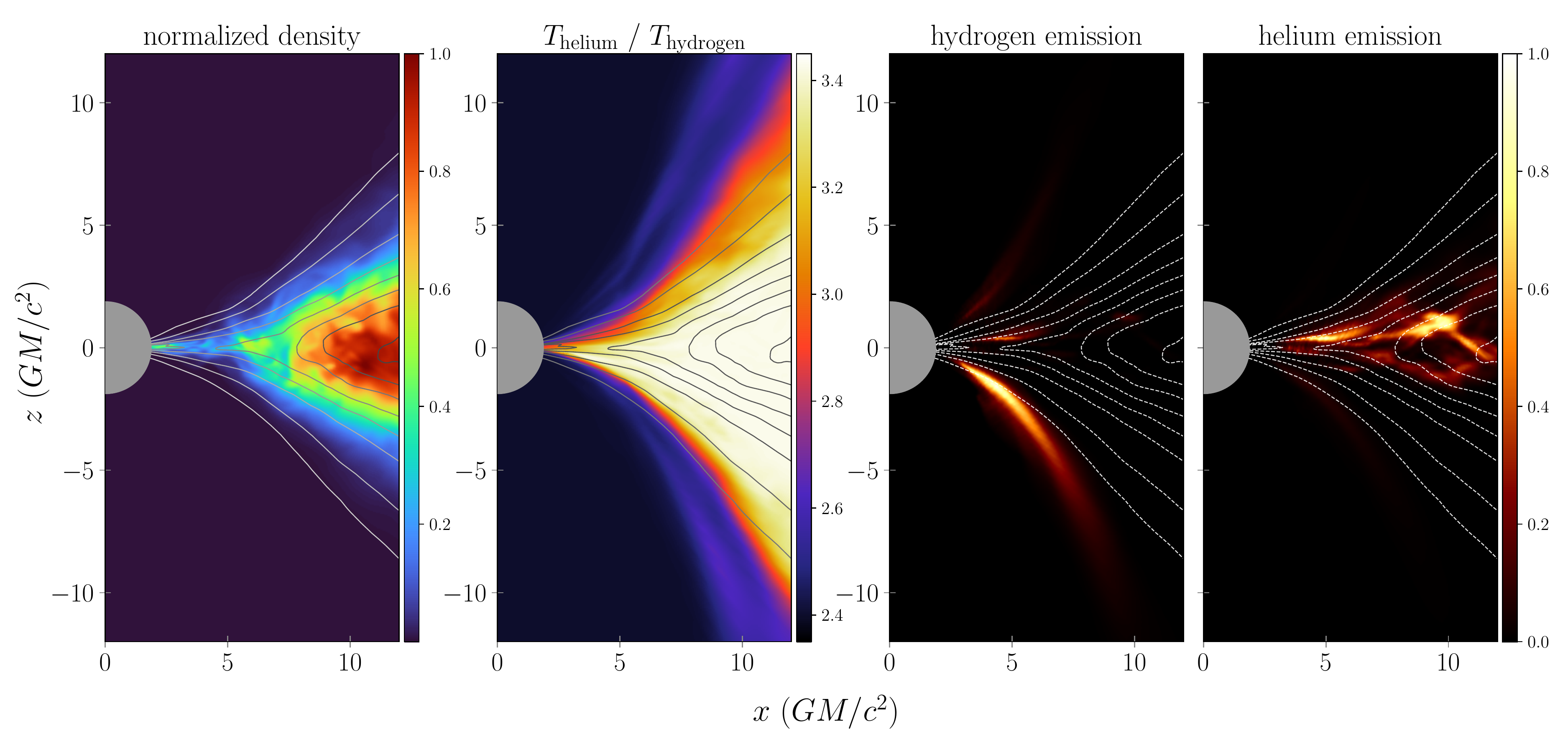}
\caption{Comparison of electron temperature and $230\,$GHz observed emission for pure-hydrogen vs.~pure-helium accretion flow. Data are for a single azimuth $\phi$ in a snapshot from a SANE $\bhspin = -0.5$ accretion simulation with $\rhigh = 10$ imaged with emission computed for an image at $17^\circ$ inclination. The left panel shows (normalized) rest-mass density of the fluid. Contours of the axisymmetrized density are overplotted on all panels to guide the eye. The second panel shows the temperature increase going from hydrogen to helium composition; regions with higher $\beta$ show a larger increase in temperature. The right two panels show the location of emission (normalized and computed as seen by the observer and accounting for the effects of optical depth) for the pure-hydrogen case and the pure-helium case. Evidently, the amount of emission along the funnel wall decreases from the hydrogen case to the helium one. All colormaps are plotted in linear scale.}
\label{fig:decompexample}
\end{figure*}

Models can be differentiated into two classes according to whether the hydrogen-composition emission peaks in the midplane or along the funnel wall. In replacing hydrogen with helium, emission tends to shift from regions of low to high $\beta$, i.e., from the funnel wall to the midplane close to the hole to the midplane farther out. The midplane typically has significantly different characteristics compared to the funnel wall, so shifting emission between these two regions has a greater effect than shifting emission within the midplane. We call the models that deviate from the one-zone model prediction \emph{funnel wall} models, as their hydrogen-composition emission peaks along the funnel wall, shifting to the midplane for helium. 

In detail, altering the gas composition at fixed $\beta$ and $\sigma$ affects the emissivity in two ways. First, changing the composition from hydrogen to helium increases the temperature of the electrons according to Equation~\ref{eqn:rhighdiff}. Since we assume $\gamma_i = 5/3$ and $\gamma_e = 4/3$, when $R = 1$ the electron temperature increases by $12/5$.\footnote{We expect the ion temperature to be larger than the electron temperature, but notice that even as $R \to 0$, a helium-only gas decreases only to twice as hot as a hydrogen-only one.} As $R \to \infty$, the factor approaches $4$. The second effect comes from the requirement that the simulations match real-world observations. Increasing $\Theta_e$ changes the emissivity of the plasma, so the other fluid parameters must change to ensure that the observed image flux density remains constant. Because $\sigma$ must remain fixed, changing the fluid mass density by a factor $\mathscr{M}$ requires changing the strength of the magnetic field by $\sqrt{\mathscr{M}}$. We typically find that the difference in absorptivities (due to the different mass density) is negligible.

The temperature increase is controlled by the local fluid parameters and thus varies across the fluid domain. In contrast, the electron number density $n_e$ and local magnetic field strength $B$ each change by a constant factor across the entire domain. Figure~\ref{fig:emissivity_contour} shows contours of the synchrotron emissivity at fixed frequency across different values of electron temperature $\Theta_e$ and local magnetic field strength. Evidently, the non-linearity of the emissivity means that the change in emissivity is non-uniform across parameter space.
These different scaling behaviors for $\Theta_e$, $B$, and $n_e$, in addition to the structure of the synchrotron emissivity, mean that the relative emissivity between different parts of the accretion flow will change with the composition. In Figure~\ref{fig:decompexample}, we show how both temperature and emission structure change for a particular funnel wall model: a SANE (retrograde) $\bhspin = -0.5$ simulation with $\rhigh = 10$. When the hydrogen gas is replaced with helium, the midplane temperature increases significantly and begins to outshine the funnel wall.

\subsection{Image statistics}

\begin{figure*}
\centering
\includegraphics[width=0.8\linewidth]{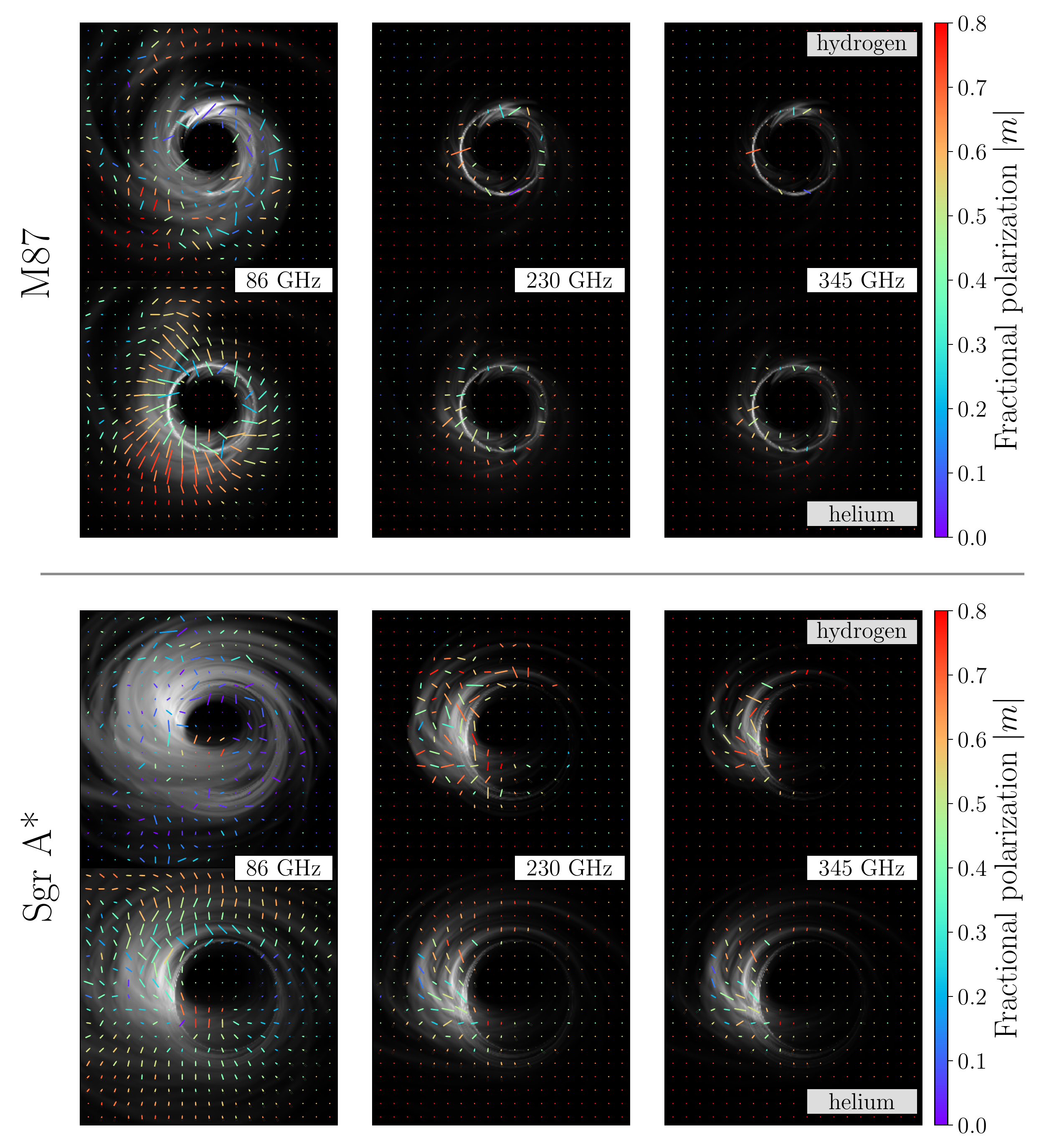}
\caption{
Comparison of polarimetric images at three frequencies for models of M87 (top) and \sgra~(bottom), with hydrogen-composition images shown above helium-composition ones. Pixel brightness denotes total intensity, the color and length of the tick marks shows linear polarization fraction / intensity, and the orientation of the ticks shows EVPA. The M87 model is from a MAD $\bhspin=0.5$ simulation with $\rhigh = 10$. The \sgra~model is SANE with $\bhspin = 0.94$, $\rhigh = 50$, and inclination $= 10^\circ$.
}
\label{fig:imgsfreqs}
\end{figure*}

We now briefly report how image properties change as a function of gas composition. We study images produced at three frequencies in the range relevant for the EHT: $86$, $230$, and $345\,$GHz. Images are produced with the \ipole~code assuming synchrotron radiation. In Figure~\ref{fig:imgsfreqs}, we show representative snapshots from both a \sgra~model and a typical M87 one, with hydrogen composition on top and helium composition below. Each panel shows the observed total intensity (grayscale) with the fractional linear polarization and EVPA plotted as the color and orientation of tick marks, respectively.
Consistent with Figure~\ref{fig:imgsblur}, the gas density is lower in the helium scenario, leading to more ordered polarization structure in the latter scenario. The image feature trends are reversed for the funnel wall models.

\begin{figure*}
\centering
\includegraphics[width=\linewidth]{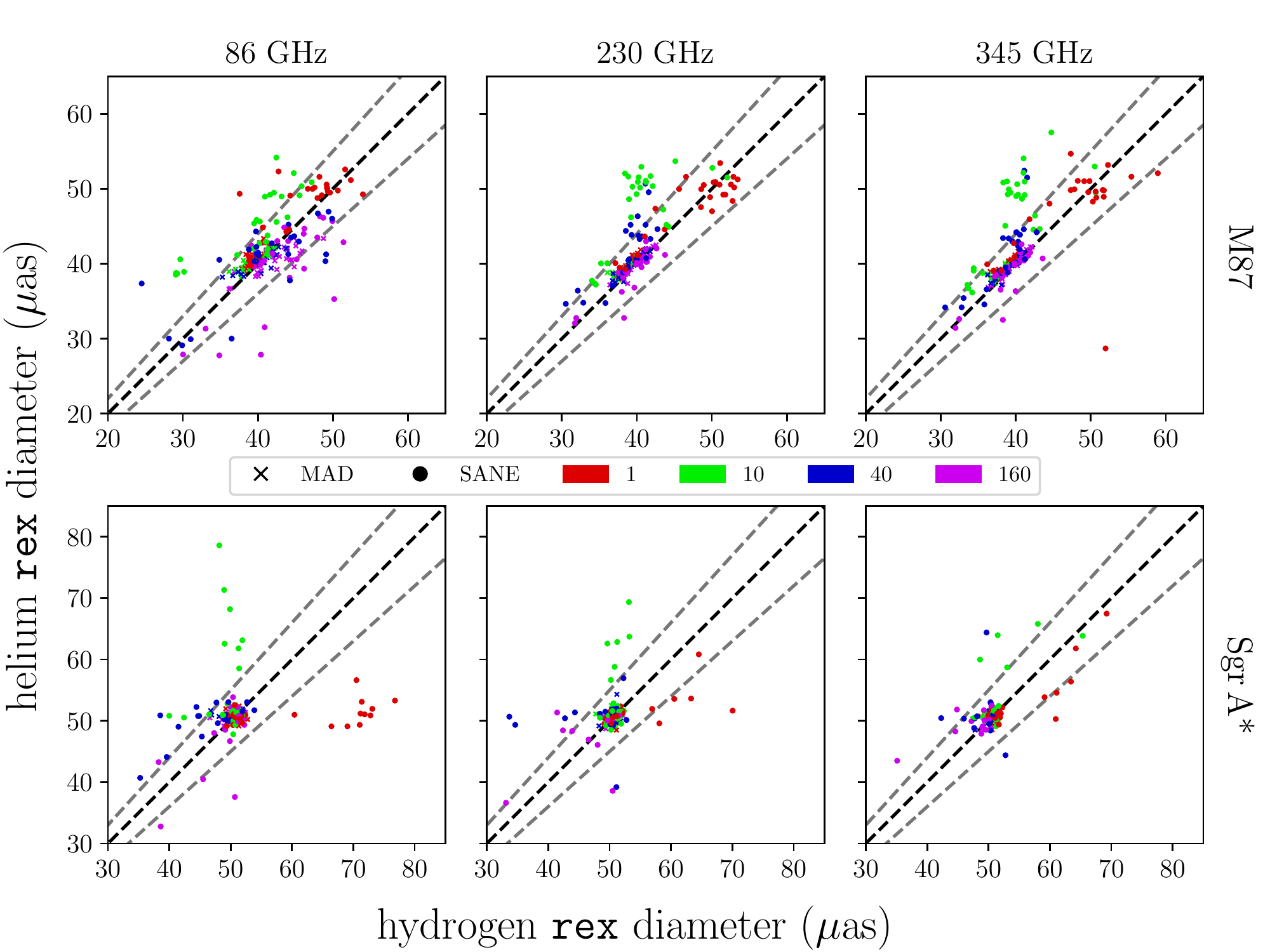}
\caption{Sample {\tt{}rex}-fit diameters for individual images from low-inclination models of M87 and \sgra~shown for pure-hydrogen composition on $x$-axis vs.~pure-helium composition on $y$-axis. Both MAD and SANE models are considered. The color of each point denotes the model $\rhigh$ and the marker type indicates the simulation magnetization state. Statistics for each model are computed for five images from the simulation movie timeseries. Points would lie on the solid black line if the measured diameter were equal for the hydrogen and helium models. Dashed lines show $\pm 10\%$ deviation.
}
\label{fig:rexdiam}
\end{figure*}

We use the ring extractor ({\tt{}rex}) method described in \S{}9 of \citet[][see also \citealt{chael_2022_ehtim}]{EHTC_2019_4} to infer ring diameters for each of our images. {\tt{}rex} performs this measurement from the image ``center point,'' which is identified as the point that is most equidistant from the peak of intensity along 360 rays cast from itself and spaced equally in angle. Figure~\ref{fig:rexdiam} shows the measured ring diameters for snapshots taken from M87 models and \sgra models that were imaged at low inclinations where a ring is easy to identify; Doppler beaming and optical depth effects often obscure the clear structure of the ring at near edge-on inclinations.

At $230$ and $345\,$GHz, the plasma is nearly optically thin and we find that the ring diameter does not change significantly. This trend is violated for SANE models with $\rhigh = 10$, when the ring diameter can increase by $\approx 20\%$. This violation occurs when emission shifts within the midplane from small radii to large radii at tens of $GM/c^2$; notice the consistency of inferred ring diameter with the SANE $\rhigh=1$ models, which have consistent extended emission at large radii. In the case of \sgra-like images, violations also arise when the hydrogen images have extended emission at large radii (as with $\rhigh = 1$), since swapping to helium composition ``thins'' the image, shifting emission toward the center and typical ``bright-ring-like'' morphology. Note that when the measured ring diameter changes drastically, the image morphology often changes significantly and becomes (in)consistent with observations.

\vspace{1em}

We also briefly explore how the polarized image properties change by considering several observables that encapsulate the net linear and circular polarizations of the image as well as the resolved linear polarization magnitude and structure.

The net linear polarization is $\mnet$ and measures the ratio of the image-integrated linear polarization $\mathcal{P} \equiv \left( \mathcal{Q}^2 + \mathcal{U}^2 \right)^{1/2}$ to the image-integrated total intensity $\mathcal{I}$, where $\mathcal{Q}$ and $\mathcal{U}$ are the image-integrated Stokes $\mathcal{Q}$ and $\mathcal{U}$ intensities. For edge-on and intermediate-inclination models (for both M87 and \sgra), we find that $\mnet$ measured for the hydrogen- and helium-composition scenarios stay within a factor of $10$ of each other, and in all cases stay below $\approx 0.1$. In fact, in most models, the change is well below a factor of $10$ and corresponds to differences in $\mnet$ of order $0.01$. In \sgra~models viewed at edge-on inclinations, $\mnet$ is typically boosted by factors of $10 - 100$. In SANE models with low values for $\rhigh$ (1, 10), linear polarization may increase drastically to as high as $\mnet \approx 0.6 - 0.7$. 

The net circular polarization $\vnet$ measures the ratio of image-integrated circular polarization magnitude $\mathcal{V}$ to $\mathcal{I}$. We find that the net circular polarization does not change significantly for any models, with M87-like models typically varying by less than a factor of $10$ and \sgra-like models varying even less, by factors of $\lesssim 5$.

The resolved linear polarization is gauged by $\avgm$, which is the ratio of the image-integrated linear polarization fraction $\mathcal{P}/\mathcal{I}$. This quantity is resolution dependent, since small coherent features in linear polarization will average out if the orientation of the linear polarization vector varies rapidly across a single resolution element (see the second and third columns of the first row of Figure~\ref{fig:imgsblur}). We find that $\avgm$ changes most drastically for SANE models, with spreads between a factor of two decrease and a factor of four increase in helium models compared to hydrogen ones. MAD models vary much less significantly. The amplification of $\avgm$ is greatest at the higher frequencies $230$ and $345\,$GHz, with the highest amplifications for $\rhigh = 1$, followed by $\rhigh = 10$ models.

\vspace{1em}

Finally, we consider how the $\beta_2$ coefficient, which measures the power in and orientation of the azimuthally symmetric mode in linear polarization vector across the image (see \citealt{palumbo_2020_DiscriminatingAccretionStates,EHTC_2021_8}). Since $\beta_2$ measures the azimuthally symmetric mode, it is often most prominent for low inclinations $\lesssim 50^\circ$, where the symmetry axis of the system can imprint on the image. We find that changes in $\beta_2$ are driven both by differences in the source structure (e.g., emission from larger radii undergo different propagation effects) and differences in local plasma properties (which will affect absorption as well as Faraday rotation and conversion).

For MAD accretion flows, we find that the magnitude of $\beta_2$ is typically large for helium gas composition compared to hydrogen, irrespective of the underlying model parameters. For SANE flows, $\left| \beta_2 \right|$ may increase or decrease, but the sense of the change is consistent with how $\avgm$ changes, which is sensible since a more coherent EVPA (required for large $\left| \beta_2 \right|$) will be related to the magnitude of the resolved linear polarization across the image. We find slight deviation from this trend in the intermediate inclination $\sim 50^\circ$ models, which often have large $\left|\beta_2\right|$ for helium compositions.

At low inclinations, the argument of $\beta_2$, which describes the average angle the linear polarization makes with a purely radial pattern, tends to be more consistent across gas composition for MAD models than SANE models. The difference is primarily because MAD models have larger $\left| \beta_2 \right|$, so that small local changes to the EVPA are less likely to scramble the overall signal. This trend is especially true for the $230$ and $345\,$GHz models. At high inclinations, the argument of $\beta_2$ changes drastically, producing a wide spread for all models.

\subsection{Variability}

We also check the effect of hydrogen vs.~helium composition on the source variability as measured in the nominal EHT band at $230\,$GHz. For this section, we use a single mass density scale to normalize all fluid snapshots from a single model so that the signal is variable and conservation laws are obeyed over the full timeseries of the lightcurve. We set the single normalization factor for each model to be the value required such that the average flux density at $230\,$GHz is $0.64\,$Jy for M87-like models or $3\,$Jy for \sgra~ones.

We report variability in terms of the modulation index 
\begin{align}
    M_{\Delta T} \equiv \dfrac{ \sigma_{\Delta T} }{ \mu_{\Delta T} },
\end{align}
where $\sigma_{\Delta T}$ is the standard deviation measured over some interval $\Delta T$ and $\mu_{\Delta T}$ is the mean measured over that same interval.
We use $M_{\Delta T}$ to be consistent with \citet{EHTC_2022_5} and because it is easy to compute and describe. In this paper, we identify $\Delta T = 3\,$hours for \sgra~$\approx 533\,GM/c^3$, which is comparable to the characteristic timescale measured in damped random walk fits to the ALMA lightcurve for the galactic center \citep{wielgus_2022_MillimeterLightCurves}. We include models for both \sgra~and M87, with $\Delta T = 533\,GM/c^3$ for both sets. Note that $533\,GM/c^3 \approx 6.5$ months for M87.

\begin{figure}
\centering
\includegraphics[width=\linewidth]{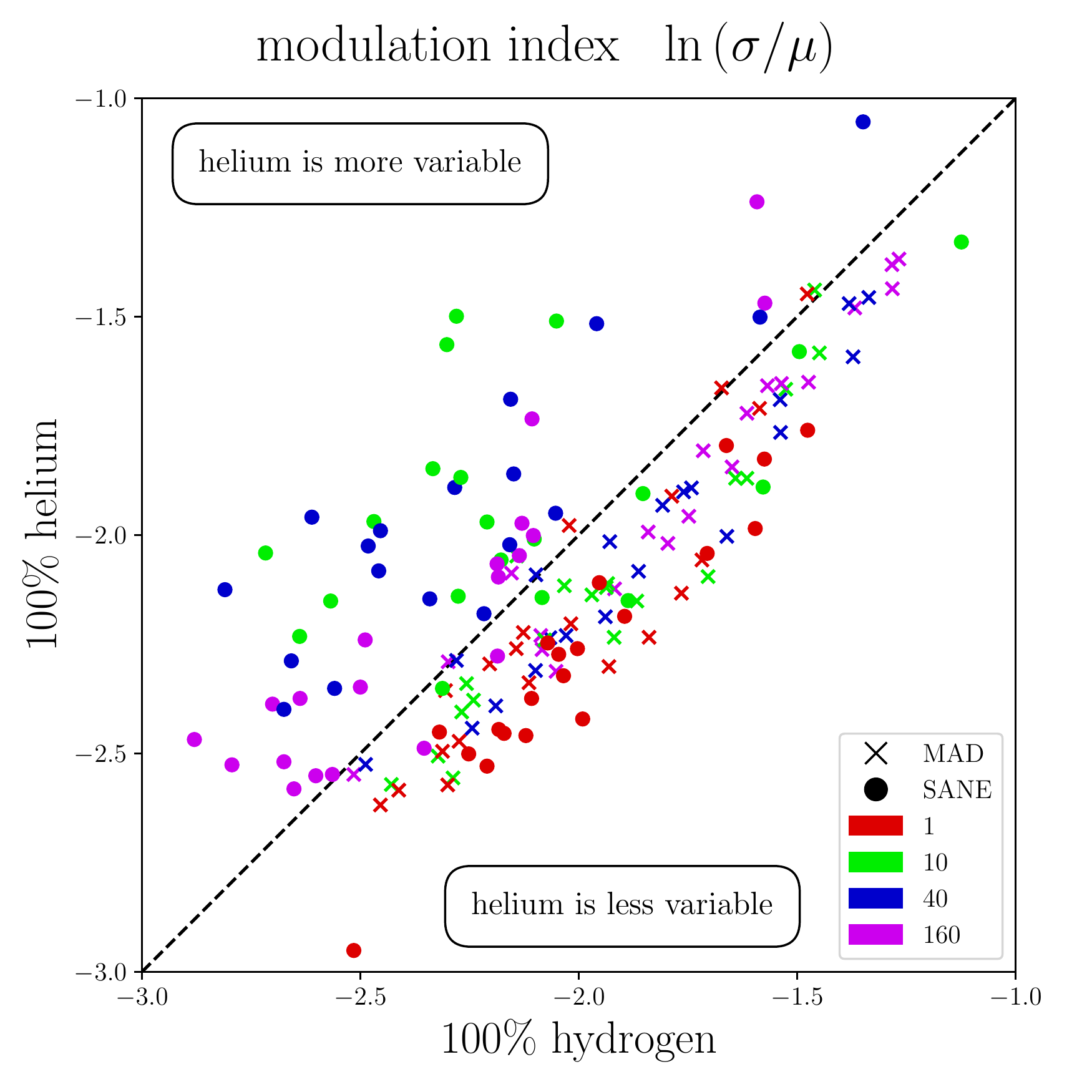}
\caption{Modulation index for hydrogen vs.~helium gas composition over all GRMHD models considered in this paper. Models of both M87 and \sgra~are included. Points in the lower-right half correspond to the case where a pure-helium gas composition is less variable than a pure-hydrogen one. Typically, MADs and SANEs with low $\rhigh$ are less variable when pure helium.}
\label{fig:variability}
\end{figure}

For every model in our catalog, we compute $M_{\Delta T}$ independently for both pure hydrogen and pure helium. Figure~\ref{fig:variability} scatter plots these data, with points lying below the diagonal line corresponding to when a pure-helium gas composition is less variable.

Evidently, helium models exhibit decreased variability when the flow is either (1) MAD or (2) SANE with a low value of $\rhigh$. This is consistent with the expectation from the emission study: in going from hydrogen to helium, emission tends to increase in regions with larger $\beta$. In the MAD models, this causes the center of emission to shift to larger radii where the evolution timescale is longer. In SANE models with low $\rhigh$, emission is primarily produced in the disk, so the center of emission similarly shifts outward to regions with longer evolution timescales. 

In contrast, for SANE models with larger $\rhigh$, a significant fraction of the emission comes from the funnel wall in the fiducial hydrogen composition scenario, and as hydrogen is replaced with helium, emission shifts toward the midplane (see, e.g., Figure~\ref{fig:decompexample}). Although more emission may also be produced at larger radii, the rapidly evolving matter near the hole imprints most strongly on the light curve, increasing the overall variability.

We caution that although gas composition influences variability by accentuating certain parts of the flow, other factors can affect observed variability as well. For example, including non-ideal effects like viscosity during the fluid evolution may alter the particular dynamics of the flow to decrease observed variability. This effect could compound or cancel out the effects of gas composition, since we have seen that the sense of variability shift due to gas composition depends on the emission source location, which is in general model dependent. Nevertheless, note that for the favored models of \sgra~identified in \citet{EHTC_2022_5}, which are MAD with $\bhspin = 0.5$ or $0.94$, $\rhigh = 160$, and at low inclination, the overall effect is to lower $M_{3\rm{}hr}$ by a factor $\approx 1.3$.  Although this reduction is not large enough to make the models consistent with the data, it is a change in the right direction.

\subsection{Spectrum statistics}

\begin{figure*}
\centering
\includegraphics[width=1\linewidth]{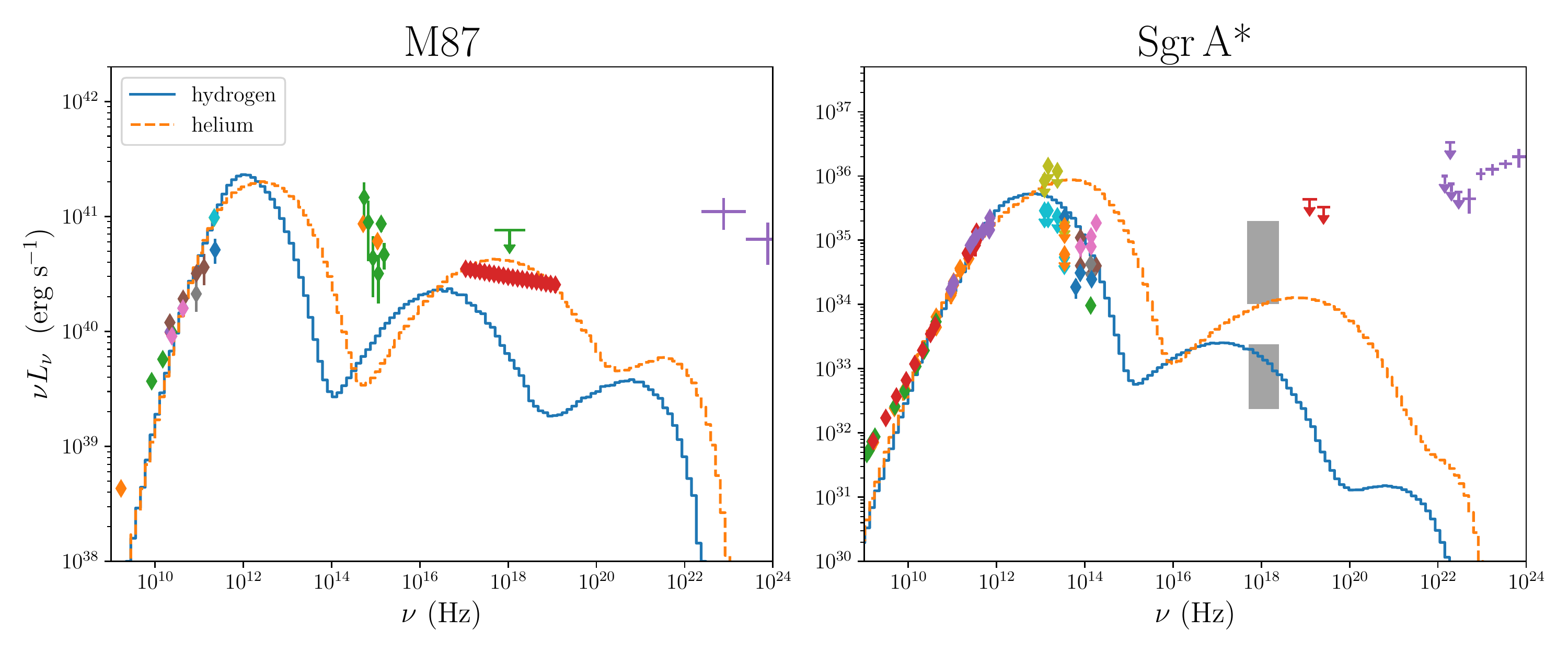}
\caption{Spectral energy distribution (SED) for hydrogen (solid) and helium (dashed) models of M87 and \sgra~plotted against observational constraints. The M87 model is a MAD with $\bhspin = 0.5$ and $\rhigh = 40$. The \sgra~model is a MAD with $\bhspin = 0.94$ and $\rhigh = 160$ at low inclination (these parameters are consistent with the best-bet models identified in \citealt{EHTC_2022_5}). The right panel (\sgra) is most representative of the average change in the SED for both sets of models. For \sgra, frequencies with multiple points correspond to the quiescent (lower) and flaring (higher) states.
References for the observational data points can be found in Appendix~\ref{app:sedrefs}.}
\label{fig:spectrum}
\end{figure*}

Finally, we explore the effect of gas composition on the model spectral energy densities (SEDs). We include the effects of synchrotron, bremsstrahlung, and Compton scattering when computing our SEDs, and we do not track polarization.\footnote{We have modified the bremsstrahlung emissivities presented in \citet{yarza_2020_BremsstrahlungGRMHDModels} for helium.} SEDs are binned over narrow ranges in inclination angle, averaged over azimuth, and mirrored across the midplane. Figure~\ref{fig:spectrum} shows typical SEDs from an M87-like model and a \sgra-like one. 

Since the local temperature of the fluid increases in our helium models, the magnitude of inverse Compton increases (see also the Compton $y$ parameter in Figure~\ref{fig:onezone}) and the SED shifts toward higher energies while obeying the constraint that it pass through the same point at the $230\,$GHz observing frequency. Typically, the magnitude of the peak also increases---the SED for \sgra~in Figure~\ref{fig:spectrum} is most representative of the typical SED behavior. Although the typical model has the helium SED lying above the hydrogen one, about $5\%$ of models (all SANE) exhibit the opposite trend, due to the geometric structure of emission in the models.

In the models we consider, the bremsstrahlung component peaks at higher frequencies than the synchrotron or synchrotron+Compton ones, at $\approx 10^{20}\,$Hz. The bremsstrahlung component in models with a helium gas composition peaks at higher frequencies than their hydrogen counterparts, but typically the overall power in the bremsstrahlung component decreases from the hydrogen to helium models. In M87-like models, the decrease is $\lesssim$ a factor of $5$ for MAD models and typically $\approx$ two orders of magnitude for SANE ones. In \sgra-like models, the SANE models also decrease by $\approx$ an order of magnitude, but the power in the MAD models' bremsstrahlung component often varies by no more than a factor of $\approx 5$ and occasionally is even greater for the helium case compared to the hydrogen one.

\section{Conclusion}
\label{sec:conclusion}

We have used analytic models and a suite of numerical simulations to study how the hydrogen--helium composition of accreting plasma can affect electromagnetic observables produced by RIAF systems. We focus on horizon-scale polarized images at $86$, $230$, and $345\,$GHz as well as spectral energy distributions (SEDs), as they are the most accessible to current and next-generation EHT-like experiments. We provide an overview of the generic results below.

In the simplest one-zone models, we assume that the ion--electron temperature ratio is held constant and find that increasing the helium fraction increases the temperature of the electrons. If the total flux density produced by the model is to match a fixed target observational value, then for the parameter ranges relevant for M87 and \sgra, the number density and magnetic field strength must decrease.

In numerical simulations, we set the ion--electron temperature ratio from the local plasma properties; this means that increasing the helium fraction increases the electron temperature by different amounts across the flow. Thus, the region of the flow that emits the most radiation may change between hydrogen and helium composition. Typically, emission shifts toward regions with higher plasma $\beta$, i.e., from the funnel wall to the disk midplane and to larger radii within the midplane itself.

We find that the inferred ring diameter in models for both M87 and \sgra~tends to vary by less than $\pm 10\%$, although in a few SANE models the image morphology changes drastically, with emission at large radii becoming more/less prominent to include (or not) a significant component of extended emission. We note that models with significantly different ring diameters have noticeably different image features, which would be detected in EHT model comparison analyses.

The net linear polarization and circular polarization fractions---integrated over the entire image---are broadly consistent between the hydrogen and helium scenarios. We find that the resolved linear polarization $\avgm$ changes most drastically when there are regions with rapidly varying EVPA for one gas composition but not the other. Which models, hydrogen or helium, have the rapidly varying EVPA is determined by the structure of the emission.

The magnitude of the $\beta_2$ observable is larger for helium composition in MAD models. In SANE models, $\left| \beta_2 \right|$ changes in line with $\avgm$, as regions with coherent EVPA produce a stronger signal in $\beta_2$. When the amplitude of $\beta_2$ is strong and the source is viewed at low inclination, we find that the argument of $\beta_2$ does not vary significantly between hydrogen and helium gas composition. 

We find that the variability of the source at $230\,$GHz decreases for MAD models and for SANE models with low $\rhigh=1$. In SANE models with larger values of $\rhigh$, variability tends to increase. The different behaviors are due to the changing emission structure: in the former set of models, emission shifts within the midplane from small radii (hydrogen) to larger radii (helium), where the characteristic times are longer; in the other models, emission shifts from the funnel wall toward the midplane close to the hole.

The spectra for helium models tend to peak at higher frequencies (in synchrotron and bremsstrahlung as well as the Compton-upscattered components). In most models the magnitude of the synchrotron and Compton spectrum also increases, although this is violated in a small fraction of SANE models. The bremsstrahlung component is typically lower in helium models, by orders of magnitude for SANE models and factors of $\approx 5$ for MAD ones.

\vspace{1em}

The analysis for the theory results published by the EHT \citep[see especially][]{EHTC_2019_5,EHTC_2021_8,EHTC_2022_5} focused heavily on accretion flows with pure hydrogen gas compositions; however, we have shown that the presence of helium can influence quantitative predictions, especially with respect to the polarimetric properties of the models. We avoid making claims about whether our results would change EHT conclusions, as the constraints applied in the EHT sequence are applied in aggregate while we consider each observable independently.

\vspace{1em}

In this work, we have limited our study to hydrogen--helium plasmas and therefore not considered the effects of heavier ions. In general, the number of ions goes as $n_i \propto X + Y/4 + Z/\left<N\right>$, where $\left<N\right>$ is the average number of nucleons per ion elements heavier than helium, and the importance of a particular species is related to both how significantly it contributes to the ion number and the ratio of ions to nucleons. Thus, given our model assumptions, we would not expect our results to change significantly as long as $Z \lesssim Y$.

In this analysis, we have assumed that the ion--electron temperature ratio is well modeled with the $\rhigh$ prescription (and fixed $\rhigh$) regardless of gas composition. The $\rhigh$ model is crude, however, in that it assumes that the temperature ratio is a function of local conditions. A more sophisticated model would  integrate an electron energy equation assuming a branching ratio that depends on local conditions for the relative dissipation into ions and electrons---for example, in the now-venerable \cite{howes_2010_PrescriptionTurbulentHeating} model for dissipation of Alfv\'{e}nic turbulence, this ratio is proportional to $(m_i/m_e)^{1/2}$. To sum up: our investigation suggests that the composition may measurably affect the appearance of EHT sources and motivates further investigation of dissipation in helium-rich plasmas.

\vspace{1em}

The authors thank Lev Arzamasskiy and Matt Kunz for thoughtful conversations and comments, as well as Michael Johnson and the anonymous referee for reading the manuscript and providing helpful feedback. GNW gratefully acknowledges support from the Institute for Advanced Study. CFG acknowledges support from NSF grants OISE 17-43747, 17-16327, and 20-34306.

\appendix

\section{Estimates of \texorpdfstring{$M/D$}{M/D} and compact flux density}
\label{app:paramestimate}

In this paper, we have considered how gas composition affects polarimetric image and SED observables for RIAF systems, which are likely descriptions of the accretion flows around \sgra~and M87. In order to synthesize images and SEDs to compare against observations of real systems, we introduce an absolute density scale and an absolute length scale to convert our numerical simulation units to physical ones. 

The density scale $\mathscr{M}$ determines the electron number density and magnetic field strength and is chosen such that the total image flux density matches an observational estimate for the compact flux density of the source at a given frequency. We find $\mathscr{M}$ with a numerical root-finding procedure.
The length scale $\mathscr{L}$ is chosen in terms of the mass of the black hole $\mathscr{L} \equiv GM_{\mathrm{BH}}/c^2$ and is determined from observations by the anticipated angular size of the hole on the sky $M/D$, where $M = M_{\mathrm{BH}}$ is the black hole mass and $D$ is the distance between the Earth and the hole. 

Both $M_{\mathrm{BH}}$ and $D$ are inputs to our models.
Here, we compile estimates for these values as well as for compact flux density at $230\,$GHz for both the M87 and \sgra~sources. Although we do choose exact numbers, we have checked and found that small deviations do not qualitatively affect our results.

\vspace{1em}

\begin{deluxetable*}{ c c c }
\tablecaption{\sgra~SED Observational Data Sources} \label{table:specrefs}
\tablehead{ 
\colhead{frequencies (Hz)} & 
\colhead{reference} &
\colhead{notes} 
}
\startdata
$43 \times 10^9 - 150 \times 10^9$ & \citet{falcke_1998_SimultaneousSpectrumSagittarius} & Table 1 \\
$1.36 \times 10^9 - 235.6 \times 10^9$ & \citet{falcke_1998_SimultaneousSpectrumSagittarius} & Table 2 \\
$3.33 \times 10^8 - 42.8 \times 10^9$ & \citet{an_2005_SimultaneousMultiwavelengthObservations} & Table 2 \\
$1.6 \times 10^9 - 353.6 \times 10^9$ & \citet{bower_2015_sgraflux} & Table 7 \\
$93 \times 10^9 - 709 \times 10^9$ & \citet{liu_2016_LinearlyPolarizedMillimeter} & Table 3 \\
$492 \times 10^9$ & \citet{liu_2016_492GHzEmission} & -- \\
$1.3 \times 10^{14}$ & \citet{eckart_1997_StellarProperMotions} & Section 6 \\
$7.97 \times 10^{13} - 1.82 \times 10^{14}$ & \citet{genzel_2003_NearinfraredFlaresAccreting} & Table 1 -- both quiescent \& flaring states \\
$1.22 \times 10^{13} - 3.45 \times 10^{13}$ & \citet{cotera_1999_MidInfraredImagingCentral} & Section 3 -- upper limits \\
$3.49 \times 10^{13}$ & \citet{schodel_2007_PossibilityDetectingSagittarius} & upper limits \\
$3.49 \times 10^{13} - 1.43 \times 10^{14}$ & \citet{schodel_2011_MeanInfraredEmission} & Table 4 \\
$1.36 \times 10^{14}$ & \citet{witzel_2012_SourceintrinsicNearinfraredProperties} & -- \\
$1.24 \times 10^{19} - 2.56 \times 10^{19}$ & \citet{goldwurm_1994_PossibleEvidenceMassive} & -- \\
$1.90 \times 10^{22} - 1.53 \times 10^{24}$ & \citet{merck_1996_StudySpectralCharacteristics} & compiled by \citet{narayan_1998_AdvectiondominatedAccretionModel} \\
\enddata
\tablecomments{
Table of references for spectral energy distribution observational constraints shown in Figure~\ref{fig:spectrum}. All frequencies are reported in Hz.
}
\end{deluxetable*}

For M87, we use the estimated value for $M/D = 3.8 \pm 0.4$ $\mu$as reported in \citet{EHTC_2019_1}. We take the distance $D = 16.8 \pm 0.8$ Mpc and the black hole mass $M_{\mathrm{BH}} = 6.5 \pm 0.7 \times 10^9\,M_\odot$, as reported in Table 1 of \citet{EHTC_2019_6}. Notice that this value differs from the gas-dynamics measurement of \citet{walsh_2013_M87BlackHole}.
We use the central estimate for compact flux density at $230\,$GHz $F_{230} = 0.64^{+0.39}_{-0.08}\,$Jy as reported in Appendix B.1 of \citet{EHTC_2019_4}.

\vspace{1em}

For \sgra, we take the black hole mass to be $M_{\mathrm{BH}}=4.1\times 10^6\,M_{\odot}$ at a distance of $D = 8.1\,$kpc, consistent with \citet{EHTC_2022_5}. \sgra~evolves on a much shorter timescale than M87, and the variability of the source makes it challenging to identify a single, representative target flux density. We choose to fit the compact flux density to $F_{230} = 3\,$Jy, but list other estimates for flux density including: $3-4\,$Jy from ALMA/SMA, as in Figure 4 of \citet{bower_2015_sgraflux}; $3\,$Jy as compiled from various in Figure 1 of \citet{dexter_2014_sgraflux}; $2.4\pm0.5\,$Jy as in \citet{doeleman_2008_ehtsgra}; and $\approx3.5\,$Jy, as in Table A2 (and text) of \citet{connors_2017_sgraflux}.

\section{Spectrum Data Sources}
\label{app:sedrefs}

Figure~\ref{fig:spectrum} in the main text plots observational data for the spectral energy distributions of M87 and \sgra. For M87, we use data compiled in \citet{ehtmwl_2021_BroadbandMultiwavelengthProperties}, which come from the EVN, HSA, VERA, EAVN, KVN, VLBA, GMVA, ALMA, SMA, HST, SWIFT, Chandra, NuSTAR, Fermi, HESS, MAGIC, VERITAS, and the EHT. We list references used to compile the SED for \sgra~in Table~\ref{table:specrefs}.

\bibliography{main.bib}
\bibliographystyle{aasjournal}

\end{document}